\documentclass[a4paper,12pt]{article}
\usepackage[pctex32]{graphics}

\begin{document}
\topmargin 0pt \oddsidemargin 0mm

\renewcommand{\thefootnote}{\fnsymbol{footnote}}
\begin{titlepage}
\vspace{5mm}
\begin{center}
{\Large \bf Entropy of black holes in topologically massive
gravity} \vspace{12mm}

{\large   Yun Soo Myung\footnote{e-mail
 address: ysmyung@inje.ac.kr}, Hyung Won Lee and Yong-Wan Kim}

\vspace{10mm} {\em  Institute of Basic Science and School of
Computer Aided Science \\ Inje University, Gimhae 621-749, Korea }
\end{center}

\vspace{5mm} \centerline{{\bf{Abstract}}}
 \vspace{5mm}

We study the issue of  black hole entropy in the topologically
massive gravity. Assuming that the presence of gravitational
Chern-Simons term with the coupling $1/\mu$ does  modify the
horizon radius $\tilde{r}_+$, we propose  $\tilde{S}_{BH}=\pi
\tilde{r}_+/2G_3$ as  the Bekenstein-Hawking entropy. This entropy
of CS-BTZ  black hole satisfies the first-law of thermodynamics
and  the area-law but it is slightly different from the
shifted-entropy $S_c=\pi r_+/2G_3+ (1/\mu l)\pi r_-/2G_3$ based on
the BTZ black hole with outer $r_+$ and inner horizon $r_-$. In
the case of $r_-=0$, $\tilde{S}_{BH}$ represents the entropy of
non-rotating BTZ black hole with the Chern-Simons term (NBTZ-CS),
while $S_c$ reduces to the entropy of NBTZ black hole. It shows
that $\tilde{S}_{BH}$ may be a candidate for the entropy of the
CS-BTZ black hole.

\end{titlepage}
\newpage
\renewcommand{\thefootnote}{\arabic{footnote}}
\setcounter{footnote}{0} \setcounter{page}{2}

\section{Introduction}

The gravitational Chern-Simons term in three dimensional Einstein
gravity produces a  propagating massive graviton~\cite{DJT}. This
theory with a negative cosmological constant $\Lambda=-1/l^2$
gives us the BTZ solution with mass $m$ and angular momentum $j$
as a trivial solution~\cite{BTZ,BHTZ}. However, there exists also
a mixed case such that ~\cite{KL,Sol}
\begin{equation} \label{mma}
M=m+\frac{j}{\mu l^2},~~J=j+\frac{m}{\mu},
\end{equation}
which shows that two conserved quantities are shifted due to the
presence of Chern-Simons term without affecting the solution of
Einstein equation. There is a chiral point at $\mu l=1$ which is a
solution to the extremal condition of $Ml=J$~\cite{LSS1}.

Also, its entropy is shifted to
\begin{equation} \label{btzc}
S_c=\frac{\pi r_+}{2G_3}+\frac{1}{\mu l}\frac{\pi r_-}{2G_3}.
\end{equation}
We observe that there is  an unusual coupling between
$\frac{1}{\mu l}$ and the inner horizon $r_-$~\cite{Park1}. This
shifted-entropy puts a difficulty on  defining its thermodynamic
relations of BTZ-CS black holes.  Importantly, it implies that the
entropy could see the region inside the outer horizon. That is,
there should be some degrees of freedom associated with the inner
horizon which would be responsible for the black hole entropy.
However, it is hard to accept this view because a black hole
entropy  relates with the outer horizon.  If an observer at
infinity describes entropy and thermodynamics of the black hole,
the region inside the outer horizon is a forbidden region.

In this Letter, we address this issue again and  propose  a new
entropy $\tilde{S}_{BH}$ which satisfies the first-law of
thermodynamics as well as the area-law of the entropy.

\section{Thermodynamics and entropy for BTZ-CS black hole}
We start with the action for the topologically mass gravity in
anti-de Sitter spacetimes~\cite{DJT}
\begin{equation}
I_{CTMG}=\frac{1}{16 \pi G_3}\int
d^3x\sqrt{-g}\Bigg[R+\frac{2}{l^2}-\frac{1}{2\mu}\varepsilon^{\lambda\mu\nu}\Gamma^\rho_{~\lambda\sigma}
\Big(\partial_{\mu}\Gamma^\sigma_{~\nu\rho}+\frac{2}{3}\Gamma^\sigma_{~\mu\tau}\Gamma^\tau_{~\nu\rho}\Big)\Bigg],
\end{equation}
where $\varepsilon$ is the tensor density defined by
$\epsilon/\sqrt{-g}$ with $\epsilon^{012}=1$. The $1/\mu$-term is
the first higher derivative correction in three dimensions because
it is the third-order derivative.

Varying  this action  leads to the Einstein equation
\begin{equation} \label{EOS}
G_{\mu\nu}+\frac{1}{\mu}C_{\mu\nu}=0,
\end{equation}
where the Einstein tensor including the cosmological constant is
given by
\begin{equation} \label{Eeom}
G_{\mu\nu}=R_{\mu\nu}-\frac{R}{2}g_{\mu\nu}
-\frac{1}{l^2}g_{\mu\nu}
\end{equation}
and the Cotton tensor is
\begin{equation}
C_{\mu\nu}=\varepsilon_\mu^{~\alpha\beta}\nabla_\alpha
\Big(R_{\beta\nu}-\frac{1}{4}g_{\beta\nu}R\Big).
\end{equation}
In the absence of Chern-Simons term, the BTZ solution to
Eq.(\ref{EOS}) is given by~\cite{BTZ}
\begin{equation}
ds^2_{BTZ}=-f(r)dt^2+\frac{dr^2}{f(r)}+r^2\Big(N^\phi
dt+d\phi\Big)^2,
\end{equation}
where $f$ and $N^\phi$ are the metric and shift functions defined
by
\begin{equation} \label{metricf}
f(r)=-8G_3m
+\frac{r^2}{l^2}+\frac{16G_3^2j^2}{r^2},~N^\phi=-\frac{4G_3j}{r^2}.
\end{equation}
Here $m$ and $j$ are the mass and the angular momentum of the BTZ
black hole, respectively. From the condition of $f=0$, we find two
horizons located at
\begin{equation} \label{hpm}
r_\pm=l\Bigg[\sqrt{2G_3\Big(m+\frac{j}{l}\Big)}\pm
\sqrt{2G_3\Big(m-\frac{j}{l}\Big)}\Bigg].
\end{equation}
We observe that in order to have  appropriate horizon radii, it
requires to have a bound of $m l\ge j$. Its thermodynamic
quantities of mass $m$, temperature $T_H=f'(r_+)/4\pi$, entropy
$S_{BH}$, angular momentum $j$, and angular velocity
$\Omega_+=-N^\phi(r_+)$ as function of $r_+$ and $r_-$ are given
by
\begin{equation}
m=\frac{r_+^2+r_-^2}{8G_3l^2},~T_H=\frac{r_+^2-r_-^2}{2 \pi
l^2r_+},~S_{BH}=\frac{\pi
r_+}{2G_3},~j=\frac{r_+r_-}{4G_3l},~\Omega_+=\frac{r_-}{lr_+}.
\end{equation}
We check that the first-law of
\begin{equation}
dm=T_H dS_{BH}+\Omega_+dj \end{equation} is satisfied and its
integral form (Smarr formula) of $m=\frac{T_HS_{BH}}{2}+\Omega_+j$
holds for the BTZ black hole. We observe the area-form of the
entropy $S_{BH}$. Here we define the on-shell condition such  that
if the first-law of thermodynamics is satisfied with thermodynamic
quantities which are obtained from using equations of motion. In
this case, the on-shell condition implies the thermal equilibrium
configuration of the BTZ black holes without any conical
singularity, while the off-shell condition means the
off-equilibrium of the BTZ black hole with conical
singularity~\cite{FFZ,BR,Myungbtz}.

 However, it is known that the Chern-Simons term does not affect
 the BTZ solution of Einstein equation because the Cotton tensor is trivially
 satisfied with the BTZ-type metric. Hence  the temperature $T_H$ and angular velocity $\Omega_+$ are fixed,
 because these are determined by solution of $f$ and $N^\phi$.

On the other hand,  the presence of Chern-Simons term has an
effect on the
 mass $M$  and angular momentum $J$ as Eq.(\ref{mma}) is shown.
These shifted-quantities were obtained using the off-shell method,
holographic gravitational anomalies~\cite{KL}. Further we note
that  the off-shell approach of conical singularity was used to
calculate the shifted-entropy $S_c$~\cite{Sol}. In addition, this
entropy was derived using other off-shell methods, Wald's Noether
formula~\cite{DSS} and Brown-Henneaux's approach~\cite{HHKT}.

Using the entropy $S_c$, it is apparent that the first-law of
black hole thermodynamics \begin{equation} \label{fir1}
 dM=T_HdS_c+\Omega_+dJ \end{equation}
  is satisfied, but
the area-law of black hole entropy violates. This implies that in
order to have the  first-law, the shift of entropy ($S_{BH}\to
S_c$) compensates changes of mass  and angular momentum ($m\to
M,j\to J$) obtained from the off-shell approach. It is noted  that
$M,S_c,J$ are off-shell quantities but $T_H$ and $\Omega_+$ are
on-shell quantities because they are derived from the on-shell
approach of metric function $f$ and $N^{\phi}$. This picture gives
rise to a difficulty in deriving  thermodynamic relations if one
considers $M(S_c,J)$ as function of $S_c$ and $J$. Actually, we
find
\begin{equation}
\Bigg(\frac{\partial M}{\partial S_c}\Bigg)_{J}\not= T_H,~
\Bigg(\frac{\partial M}{\partial J}\Bigg)_{S_c}\not= \Omega_+.
\end{equation}
Furthermore, we have a difficulty in defining the heat capacity,
which is an important thermodynamic quantity to be used for
testing the thermodynamic stability. Its usual definition
\begin{equation} \label{heat}
 C_{J}=\Bigg(\frac{\partial M}{\partial T_H}\Bigg)_J  \end{equation}
does not work because it is defined by $M$(off-shell quantity) and
$T_H$(on-shell quantity).
 In order to cure this problem, we reexpress
Eq.(\ref{fir1}) as
\begin{equation} \label{first2}
 dM=T_HdS_{BH}+\Big(\Omega_++\frac{1}{\mu l^2}\Big)dj \end{equation}
 which shows that the change of angular velocity compensates the
 change of mass.
Here we can check the thermodynamic relations
\begin{equation}
\Bigg(\frac{\partial M}{\partial S_{BH}}\Bigg)_{j}= T_H,~
\Bigg(\frac{\partial M}{\partial j}\Bigg)_{S_{BH}}=
\Omega_++\frac{1}{\mu l^2}
\end{equation}
since we can express the shifted-mass as function of $S_{BH}$ and
$j$
\begin{equation}
M(S_{BH},j)=\frac{G_3S^2_{BH}}{2\pi^2l^2}+\frac{\pi^2j^2}{2G_3S_{BH}^2}+\frac{j}{\mu
l^2}.
\end{equation}
 In this case, we can define the heat capacity as that of the BTZ
 black hole
\begin{equation} \label{heat2}
 C_{j}=\Bigg(\frac{\partial M}{\partial T_H}\Bigg)_j=\frac{\pi
r_+\Delta}{2G_3(2-\Delta)}=\Bigg(\frac{\partial m}{\partial
T_H}\Bigg)_j  \end{equation} with
$\Delta=\sqrt{1-\frac{j^2}{m^2l^2}}$. This is independent of the
CS-coupling $\mu$.
 It implies  that the shifted-quantities of $M,J,S_c$ are not suitable for computing
thermodynamic relations because all they are mixed as functions of
$(r_+,r_-)$.

Let us discuss what happens for the extremal condition of
$T_H=0=C_j$ when using the shifted-quantities. In the absence of
the Chern-Simons term, the extremal black hole is located at
$r_+=r_-=r_e$ for $ml=j$. In the presence of the Chern-Simons
term, this condition is $Ml=J$, which implies
\begin{equation} \label{chiralc}
\mu l=1,~ml=j, \end{equation} where the first one is new for  the
BTZ-CS black hole  and is called a chiral point. However, there is
no way to implement this chiral condition (extremal black hole)
without changing  the temperature $T_H$ and heat capacity $C_j$
(metric function $f$).

\section{Thermodynamics and entropy for CS-BTZ black hole}
We are in a position to seek other solution by  implementing the
shifted condition of Eq. (\ref{mma}). We suggest that  this
condition changes the form of metric function. Plugging this into
Eq.(\ref{metricf}) ($m \to M,j\to J$), we find two
shifted-horizons for the CS-BTZ located at
\begin{equation} \label{shpm}
\tilde{r}_\pm=l\Bigg[\sqrt{1+\frac{1}{\mu
l}}\sqrt{2G_3\Big(m+\frac{j}{l}\Big)}\pm \sqrt{1-\frac{1}{\mu
l}}\sqrt{2G_3\Big(m-\frac{j}{l}\Big)}\Bigg]
\end{equation}
which shows clearly that the presence of Chern-Simons term
modifies the black hole horizons. In this process, the Einstein
equation (\ref{Eeom}) is trivially satisfied. This is our main
result which differs from the BTZ-CS. We mention the difference
between CS-BTZ and BTZ-CS. We mean by CS-BTZ that the Chern-Simons
term shifts the horizon radii through Eq.(\ref{mma}). On the other
hand, the BTZ-CS implies that the horizon radii remain unchanged,
while the mass and angular momentum are shifted so that the
shifted-entropy does appear.

Fortunately, we find that the degenerate horizon of
\begin{equation}
\tilde{r}_+=\tilde{r}_-\equiv \tilde{r}_e \end{equation} which
provides  Eq.(\ref{chiralc}) of  $\mu l=1$ and $ lm=j$. The chiral
point of $\mu l=1$~\cite{LSS1,CDWW} corresponds to the extremal
CS-BTZ at $\tilde{r}_e=2l\sqrt{G_3(m+j/l)}$.
 From the condition of
$\tilde{r}_\pm \ge 0$, we have the bound for $\mu$, in addition to
$lm \ge j$:
\begin{equation}
\mu l \ge 1.
\end{equation}

 For the CS-BTZ black holes, we could define the
thermodynamic quantities of mass $M$, new temperature
$\tilde{T}_H$,
 angular momentum $J$, and new angular velocity $\tilde{\Omega}_+$
as function of $\tilde{r}_+$ and $\tilde{r}_-$,
\begin{equation}
M=\frac{\tilde{r}_+^2+\tilde{r}_-^2}{8G_3l^2},~\tilde{T}_H=\frac{\tilde{r}_+^2-\tilde{r}_-^2}{2
\pi
l^2\tilde{r}_+},~J=\frac{\tilde{r}_+\tilde{r}_-}{4G_3l},~\tilde{\Omega}_+=\frac{\tilde{r}_-}{l\tilde{r}_+}.
\end{equation}
Here we note that the shifted-mass $M$ and angular momentum $J$
are expressed in a compact way when using $\tilde{r}_+$ and
$\tilde{r}_-$. For $\tilde{r}_+=\tilde{r}_-$,  we find the thermal
condition for the extremal black hole
\begin{equation}
\tilde{T}_H=0,~\tilde{C}_J=0, \end{equation} where the new heat
capacity is given by
\begin{equation}
\tilde{C}_{J}=\frac{\pi
\tilde{r}_+\tilde{\Delta}}{2G_3(2-\tilde{\Delta})} \end{equation}
with
\begin{equation}
\tilde{\Delta}=\sqrt{1-\frac{J^2}{M^2l^2}}=\sqrt{1-\frac{1}{\mu^2l^2}}\sqrt{1-\frac{j^2}{m^2l^2}}.
\end{equation}

 Importantly, the new entropy is given  by the Bekenstein-Hawking
entropy as
\begin{equation} \label{nent}
\tilde{S}_{BH}=\frac{\pi \tilde{r}_+}{2G_3}=\frac{\pi
l}{2G_3}\Bigg[\sqrt{1+\frac{1}{\mu
l}}\sqrt{2G_3\Big(m+\frac{j}{l}\Big)}+ \sqrt{1-\frac{1}{\mu
l}}\sqrt{2G_3\Big(m-\frac{j}{l}\Big)}\Bigg]
\end{equation}
which is slightly different from the  shifted-entropy $S_c$ of
Eq.(\ref{btzc}) expressed as~\cite{SS,SST,DSS}
\begin{equation}\label{cent}
S_c=\frac{\pi l}{2G_3}\Bigg[\Big(1+\frac{1}{\mu
l}\Big)\sqrt{2G_3\Big(m+\frac{j}{l}\Big)}+ \Big(1-\frac{1}{\mu
l}\Big)\sqrt{2G_3\Big(m-\frac{j}{l}\Big)}\Bigg].
\end{equation}
Here we check easily  that the first-law of thermodynamics
\begin{equation}
dM=\tilde{T}_Hd\tilde{S}_{BH}+\tilde{\Omega}_+dJ
\end{equation}
is satisfied for the CS-BTZ black hole.
 This implies that new
thermodynamic quantities  (tilde variables) compensate the shifts
of mass and angular momentum to have the first-law and the
area-law form of the entropy.  Furthermore, we confirm that
thermodynamic relations hold for the CS-BTZ black hole:
\begin{equation}
\Bigg(\frac{\partial M}{\partial \tilde{S}_{BH}}\Bigg)_{J}=
\tilde{T}_H,~ \Bigg(\frac{\partial M}{\partial
J}\Bigg)_{\tilde{S}_{BH}}= \tilde{\Omega}_+.
\end{equation} Especially, we have  the Smarr
formula of
$M=\frac{\tilde{T}_H\tilde{S}_{BH}}{2}+\tilde{\Omega}_+J$.
 In
order for $\tilde{S}_{BH}$ to compare with $S_c$, we make a
 series expansion for $\mu l \gg 1$ as
 \begin{equation}
 \tilde{S}_{BH} \simeq \frac{\pi r_+}{2G_3}+\frac{1}{2\mu l}\frac{\pi r_-}{2G_3}
\end{equation}
where the last term differs slightly from that of $S_c$.

 At this stage, we introduce two similar cases in higher dimensions.
Let us observe what happens for the Reissner-Nordstr\"om-AdS black
holes when the Gauss-Bonnet term with coupling constant $a$ is
added~\cite{MS,ABD}. Its action $S_d$ is given by $16\pi G_d
S_d=\int d^dx(R-2\Lambda-F^2/4+a{\cal L}_{GB})$. For $d=4$, using
the entropy function approach, one finds that
$S_{4}=(\pi/G_4)(v_2+4a)$, which means that the Gauss-Bonnet term
does not modify the horizon radius ($v_2$ is independent of $a$)
but it shifts the black hole entropy. Here $v_2$ is the radius
square appeared in $ds^2=v_1(-r^2dt^2+dr^2/r^2)+v_2d\Omega_{d-2}$
of $AdS_2\times S^{d-2}$. In this case, there is no change of
equations of motion even for the presence of Gauss-Bonnet term
because the Gauss-Bonnet term plays the role of a topological
term. This is the same situation as  the Chern-Simons term is
added in three dimensions.  However, for $d=5$, we have
$S_{5}=(\pi^2\sqrt{v_2}/2G_5)(v_2+12a)$. Here $v_2$ is a function
of $a$. The $a$-dependent term represents the deviation from the
area-law due to the Gauss-Bonnet term.  In this case, there is
change of equations of motion in the presence of Gauss-Bonnet
term.  Hence, our entropy $\tilde{S}_{BH}$ is between $S_{4}$ and
$S_5$ because the horizon radius was changed from $r_+$ to
$\tilde{r}_+$.

On the other hand, from the CFT on the boundary at infinity, we
can calculate the entropy using the Cardy-formula. Taking into
account the shifted mass and angular momentum in Eq.(\ref{mma}),
the shifted Virasoro generators are given by~\cite{LSS1,HHKT}
\begin{equation} \label{sva}
\tilde{L}^{\pm}_0=\frac{l}{2}\Big(M\pm\frac{J}{l}\Big)=
\frac{l}{2}\Big(1\pm\frac{1}{\mu l}\Big)\Big(m\ \pm
\frac{j}{l}\Big).
\end{equation}
Introducing
 the Cardy formula
 \begin{equation}
 S^{CFT}= 2\pi \sqrt{\frac{c_L \tilde{L}_0^+}{6}}+2\pi \sqrt{\frac{c_R
 \tilde{L}_0^-}{6}},
 \end{equation}
we recover $\tilde{S}_{BH}$ by choosing
\begin{equation}
\tilde{c}_{L/R}=\frac{3l}{2G_3},
\end{equation}
while  $S_c$  is obtained   by choosing  the shifted central
charges
\begin{equation} \label{fcc}
c_{L/R}=\frac{3l}{2G_3}\Big(1\pm \frac{1}{\mu l}\Big).
\end{equation}

\section{Thermodynamics and entropy for CS-NBTZ black holes}
We are in a position to consider the  mixed case with $j=0$ in
Eq.(\ref{mma}) which  is equivalent  to the BTZ black hole with
$m=m$ and angular momentum $j=m/\mu$. In this case, we denote
\begin{equation}
M^{j=0}=m,~~J^{j=0}=\frac{m}{\mu},
\end{equation}
which  indicates  that the presence of Chern-Simons term shifts
the angular momentum but does not change the mass if we start with
$j=0$ solution. Hence, the $j=0$ case is more attractive to show
the effect of the Chern-Simons term than the $j\not=0$ case.
Plugging this into Eq.(\ref{hpm}), we find two shifted-horizons
for the CS-NBTZ located at
\begin{equation} \label{hpmz}
\tilde{r}^{j=0}_\pm=l\sqrt{2G_3m}\Bigg[\sqrt{1+\frac{1}{\mu l}}\pm
\sqrt{1-\frac{1}{\mu l}}\Bigg]
\end{equation}
which shows clearly that the presence of Chern-Simons terms
modifies the black hole horizons. Explicitly, the presence of
Chern-Simons term creates a new inner horizon $\tilde{r}_-$, in
compared with the single horizon of NBTZ black hole. Furthermore,
at the chiral point of $\mu l=1$,  we have a new  degenerate
horizon at $\tilde{r}_e^{j=0}=2l\sqrt{G_3m}$.

For the CS-NBTZ, we define the thermodynamic quantities of mass,
temperature, angular momentum, and angular velocity as function of
$\tilde{r}_+$ and $\tilde{r}_-$ in Eq.(\ref{hpmz}),
\begin{equation}
M^{j=0}=\frac{\tilde{r}_+^2+\tilde{r}_-^2}{8G_3l^2},~T_H^{j=0}=\frac{\tilde{r}_+^2-\tilde{r}_-^2}{2
\pi
l^2\tilde{r}_+},~J^{j=0}=\frac{2\tilde{r}_+\tilde{r}_-}{8G_3l},~\Omega_+^{j=0}=\frac{\tilde{r}_-}{l\tilde{r}_+}.
\end{equation}
Crucially, the entropy is given by the Bekenstein-Hawking formula
as
\begin{equation} \label{nentp}
\tilde{S}_{BH}^{j=0}=\frac{\pi \tilde{r}^{j=0}_+}{2G_3}=\pi l
\sqrt{\frac{m}{2G_3}}\Bigg[\sqrt{1+\frac{1}{\mu l}}+
\sqrt{1-\frac{1}{\mu l}}\Bigg]
\end{equation}
which contains the effect of Chern-Simons term. For the critical
point of  $\mu l=1$, one has the extremal entropy
$\tilde{S}_{BH}^{j=0}=\pi l \sqrt{m/G_3}$, while for $\mu l \to
\infty$, one finds $\tilde{S}_{BH}^{j=0}=\pi l \sqrt{2m/G_3}$
which is just the entropy of the NBTZ black hole.

 On the other hand, the shifted-entropy $S_c^{j=0}$ is just the entropy of NTBZ black hole as
\begin{equation}\label{centp}
S_c^{j=0}=\pi l \sqrt{\frac{2m}{G_3}}=S^{NBTZ},
\end{equation}
where shows clearly  that there is no effect of Chern-Simons terms
on the black hole entropy for $j=0$ case, as is implied from
Eq.(\ref{btzc}) with $r_-=0$. Furthermore, there is no chiral
point of $\mu l=1$. This means that there is no shift of
Chern-Simons term to the entropy when using the NBTZ background.

\section{Discussions}
\begin{table}
 \caption{Summary of thermodynamic picture
 for BTZ, BTZ-CS and CS-BTZ. Here First-L, TR, Heat-C, and  Smarr-F represent the first-law of thermodynamics, thermodynamic relations, heat capacity, and  Smarr formula, respectively.}
 \begin{tabular}{|c|c|c|c|}
 \hline
  & BTZ  & BTZ-CS & CS-BTZ \\ \hline
  First-L &$dm=T_H dS_{BH}+\Omega_+dj$& $dM=T_HdS_c+\Omega_+dJ$& $dM=\tilde{T}_Hd\tilde{S}_{BH}+\tilde{\Omega}_+dJ$ \\ \hline
  TR & $\Big(\frac{\partial M}{\partial S_{BH}}\Big)= T_H,
\Big(\frac{\partial M}{\partial j}\Big)= \Omega_+$& N/A&
$\Big(\frac{\partial M}{\partial \tilde{S}_{BH}}\Big)=
\tilde{T}_H, \Big(\frac{\partial M}{\partial J}\Big)=
\tilde{\Omega}_+$ \\ \hline Entropy &$S_{BH}=\frac{\pi
r_+}{2G_3}$& $S_c=\frac{\pi r_+}{2G_3}+\frac{1}{\mu l}\frac{\pi
r_-}{2G_3}$& $S_{BH}=\frac{\pi \tilde{r}_+}{2G_3}$\\ \hline Heat-C
&$C_{j}=\frac{\pi r_+\Delta}{2G_3(2-\Delta)}$& N/A&
$\tilde{C}_{J}=\frac{\pi
\tilde{r}_+\tilde{\Delta}}{2G_3(2-\tilde{\Delta})}$\\
\hline
 Smarr-F &$m=T_HS_{BH}/2+\Omega_+j$& $M=T_H S_{c}/2+\Omega_+J$& $M=\tilde{T}_H\tilde{S}_{BH}/2+\tilde{\Omega}_+J$ \\ \hline

 \end{tabular}
 \end{table}
First of all, we summarize our thermodynamic results for BTZ,
BTZ-CS, and CS-BTZ black holes  in Table 1 for comparison.

We recover the shifted-mass $M$,  angular momentum $J$ and the
chiral condition of $\mu l=1$ from shifted-horizons of the CS-BTZ
black hole. In this case, we obtain a new entropy
$\tilde{S}_{BH}$.  Also the consistent thermodynamic relations are
derived. This is a good achievement obtained  by plugging the
shifted-mass and angular momentum into the metric function $f$ and
$N^\phi$.

Although the shifted-entropy $S_c$ is well-known as the landmark
of the topologically massive gravity, it puts a difficulty on
defining  thermodynamic relations of BTZ-CS black holes.
Importantly, it implies that the entropy could see the region
inside the outer horizon because of the coupling between $1/\mu$
and the inner horizon $r_-$. However, it is hard to accept this
view because a black hole entropy relates with the outer horizon.

The  entropy of $\tilde{S}_{BH}$  seems to be a candidate for the
entropy for the CS-BTZ black hole  since it satisfies the
first-law of thermodynamics and the area-law of the black hole
entropy. The effect of the Chern-Simons term shifts the horizons
and creates a new degenerate horizon at the chiral point of $\mu
l=1$ even for the NBTZ background of $m\not=0, j=0$. The
 point  to note is the equivalence between the CS-NBTZ and the
BTZ black hole, which shows the role of the Chern-Simons term in
the black hole clearly.

However, one thing to be understood is why the entropy of
$\tilde{S}_{BH}$ could be recovered from the CFT on the boundary
at infinity with the central charges of $\tilde{c}_{R/L}=3l/2G_3$.
Assuming the AdS$_3$/CFT$_2$ correspondence, the bulk  theory of
topologically massive gravity is equivalent to a conformal field
theory with different right and left central charges $c_{R/L}$  in
Eq. (\ref{fcc}). In order to recover the shifted-entropy $S_c$,
one has to take into account the shifted-Virasoro generators in
Eq. (\ref{sva}) as well as the shifted-central charges $c_{R/L}$.
This may correspond to  doubly counting the Chern-Simons terms for
computation of the entropy. In deriving the entropy
$\tilde{S}_{BH}$, we did not use the central charges $c_{R/L}$.

\section*{Acknowledgment}
Y. Myung was  supported by the Science Research Center Program of
the Korea Science and Engineering Foundation (KOSEF) through the
Center for Quantum Spacetime of Sogang University with grant
number R11-2005-021. H. Lee was  in part supported by KOSEF,
Astrophysical Research Center for the Structure and Evolution of
the Cosmos at Sejong University.

        \end{document}